\documentclass[conference,a4paper]{IEEEtran}
\IEEEoverridecommandlockouts

\usepackage[hidelinks]{hyperref}
\usepackage[cmex10]{amsmath}
\usepackage{amssymb,amsfonts}
\usepackage{dblfloatfix}

\usepackage[ruled,vlined]{algorithm2e}
\usepackage{graphicx}
\graphicspath{{Figures/PDF/}{Figures/PNG/}{Figures/BMP/}}

\usepackage{booktabs}
\usepackage{siunitx}
\usepackage[numbers,compress]{natbib}
\usepackage{texnames}
\usepackage{bm,bbm}
\usepackage{orcidlink}
\usepackage{subcaption}

\begin{document}

\title{\uppercase{When marine radar target detection meets pretrained large language models}
\thanks{This work has been accepted for publication in the Proceedings of the IEEE International Geoscience and Remote Sensing Symposium (IGARSS 2025).}
\thanks{Corresponding author: Xueqian Wang(wangxueqian@mail.tsinghua.edu.cn).}}

\author{
    Qiying Hu$^{1}$, Linping Zhang$^{1}$, Xueqian Wang$^{1,2}$, Gang Li$^{1,2}$, Yu Liu$^{1}$, Xiao-Ping Zhang$^{3}$\\
    $^{1}$Department of Electronic Engineering, Tsinghua University, Beijing, China \\
    $^{2}$State Key Laboratory of Space Network and Communications, Tsinghua University, Beijing, China\\
    $^{3}$Shenzhen Key Laboratory of Ubiquitous Data Enabling, Tsinghua Shenzhen International\\ Graduate School, Tsinghua University, Shenzhen, China
}

\maketitle

\begin{abstract}
Deep learning (DL) methods are widely used to extract high-dimensional patterns from the sequence features of  radar echo signals. However, conventional DL algorithms face challenges such as redundant feature segments, and constraints from restricted model sizes. To address these issues, we propose a framework that integrates feature preprocessing with large language models (LLMs). Our preprocessing module tokenizes radar sequence features, applies a patch selection algorithm to filter out uninformative segments, and projects the selected patches into embeddings compatible with the feature space of pre-trained LLMs. Leveraging these refined embeddings, we incorporate a pre-trained LLM, fine-tuning only the normalization layers to reduce training burdens while enhancing performance. Experiments on measured datasets demonstrate that the proposed method significantly outperforms the state-of-the-art baselines on supervised learning tests.
\end{abstract}

\begin{IEEEkeywords}
Marine target detection, large language models (LLMs), patch selection
\end{IEEEkeywords}

\section{Introduction}
Target detection in the presence of sea clutter has long been a critical and challenging problem in radar target detection. Recently, approaches leveraging multi-domain features of radar echoes have garnered significant attention, utilizing phase, Doppler, and time-frequency domain characteristics to distinguish targets from sea clutter. Several manually crafted methods~\cite{shui2014tri,shi2018sea,xie2021phase,zhao2021eigenvalues} have been developed to extract statistical properties from these multi-domain features. However, these methods rely heavily on domain expertise and handcrafted heuristics, often struggling to capture high-dimensional patterns in the signal data.

With the rapid development of deep learning technology, convolutional neural networks (CNNs) for time-frequency feature extraction and long short-term memory networks (LSTM) for sequence feature extraction have been adopted in radar target detection. Chen \textit{et al.}~\cite{chen2021false} design a dual-channel CNN (DCCNN)-based structure detector that extracts both amplitude and time-frequency information from signals to achieve target detection. Qu \textit{et al.}~\cite{qu2023enhanced} introduce an attention-enhanced CNN to capture and learn the deep features of Wigner–Ville distribution of radar signal. Wan \textit{et al.}~\cite{wan2022sequence} propose a sequence feature-based detector based on instantaneous phase feature, Doppler spectrum feature, short-time Fourier transform feature, and bidirectional long short-term memory network (Bi-LSTM). 

Despite advancements in deep learning-based detectors, several limitations hinder their practical applications. A major challenge is the presence of redundant and irrelevant segments in radar signal features~\cite{riti2023feature}, which may degrade detection performance. Inspired by~\cite{lin2024rho,zhang2024sparsevlm}, which highlight that not all text and image tokens are necessary for training, we propose a patching and patch selection strategy that filters out irrelevant information in radar signal features, thereby enhancing model performance. Another critical limitation stems from the constrained capacity of small models. Recent studies~\cite{zhou2023one,liu2024unitime,jin2023time,yuan2024unist,zheng2024large} highlight the exceptional cross-modal transfer capabilities of pre-trained large language models (LLMs). Despite being trained on textual data, LLMs exhibit remarkable generalization, extending their feature recognition abilities to time-series modalities. Solid analyses~\cite{zhou2023one} further reveal that the self-attention mechanism in LLMs operates analogously to Principal Component Analysis (PCA), enabling the extraction of key components from high-dimensional data. This insight opens a very promising direction for radar target detection by leveraging pre-trained LLMs to replace traditional small models, offering the potential for significant performance enhancements.

\begin{figure*}[t]
\centering
    \includegraphics[width=0.95\textwidth]{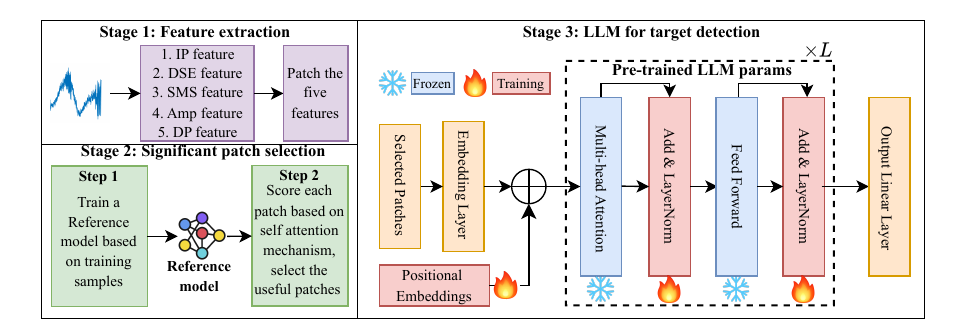}
    \caption{Overview of our LLM-empowered target detection method.}
    \label{RadarGPT}
\end{figure*}

In this paper, we present a novel approach for marine radar target detection powered by LLMs. Our methodology is outlined in Fig.~\ref{RadarGPT}. Initially, we extract five sequence features from radar echo signals and segment them into multiple feature patches. We then employ a reference model to score each patch, identifying the most relevant ones for target detection. Finally, we develop a target detection model based on the pre-trained transformer architecture GPT-2~\cite{radford2019language}. Through fine-tuning, our method improves the average detection rate by 18.19\% compared to a recent sequence feature-based approach~\cite{wan2022sequence} and surpasses a state-of-the-art method~\cite{qu2023enhanced} by 5.88\% across different real-world datasets.

\section{Proposed method}
\subsection{Signal feature extraction} \label{SF}
When the radar transmits coherent pulses toward the sea, it receives a time series of echoes for each distance cell. In clutter cells, echoes consist of sea surface scatter and noise, while in target cells, they include target echoes, sea surface scatter, and noise. Target detection is thus a binary hypothesis test to determine if the echo contains a target component. The time series of echoes $x$ from unit can be split into observation vectors ${x}_i$ in terms of:
\begin{equation} \label{Pr}
    x_i=[x_{M\cdot (i-1)+m}]_{m=1}^{N},\quad i=1,2,...
\end{equation}
where $N$ denotes the length of the observation, and $M$ denotes the interval length of the observation vectors. Following the methods outlined in Ref.~\cite{wan2022sequence} and Ref.~\cite{wang2023multi}, we adopted five sequence features from the observation: Instantaneous Phase (IP), Doppler Spectral Entropy (DSE), STFT Marginal Spectrum (SMS), Amplitude (Amp), and Doppler Phase (DP). To extract local semantic information, we utilize patching~\cite{nietime} by aggregating
adjacent time steps to form a single patch-based token. Specifically, the IP feature $F_{\mathrm{IP}}$, DSE feature $F_{\mathrm{DSE}}$, SMS feature $F_{\mathrm{SMS}}$, Amp feature $F_{\mathrm{Amp}}$, and DP features $F_\mathrm{DP}$ are combined to form the input feature matrix ${F}=[F_{\mathrm{IP}};F_{\mathrm{DSE}};F_{\mathrm{SMS}};F_{\mathrm{Amp}};F_\mathrm{DP}]\in \mathbb{R}^{5\times N}$. Each feature in ${F}$ is then partitioned into non-overlapping segments of length $L$, zero-padding is applied to the last patch that is not fully filled. These $K$ segments are concatenated together to form the final input ${F}^{\mathrm{P}}\in \mathbb{R}^{K\times L}$, where $K=5\left\lceil\frac{N}{L}\right\rceil$.

\begin{figure}[h] 
\centering    
\includegraphics[width=0.5\textwidth]{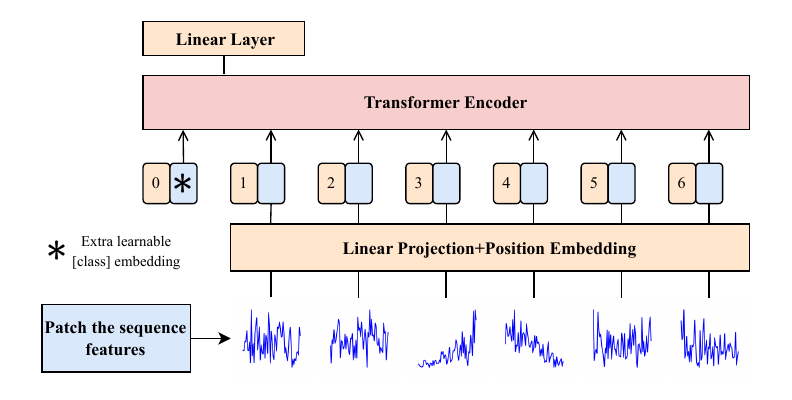}    
\caption{Overview of the reference model.} \label{RE}
\end{figure}

\begin{figure}[htb]
\centering
\begin{minipage}[b]{0.49\linewidth}
  \centering
  \includegraphics[width=\linewidth]{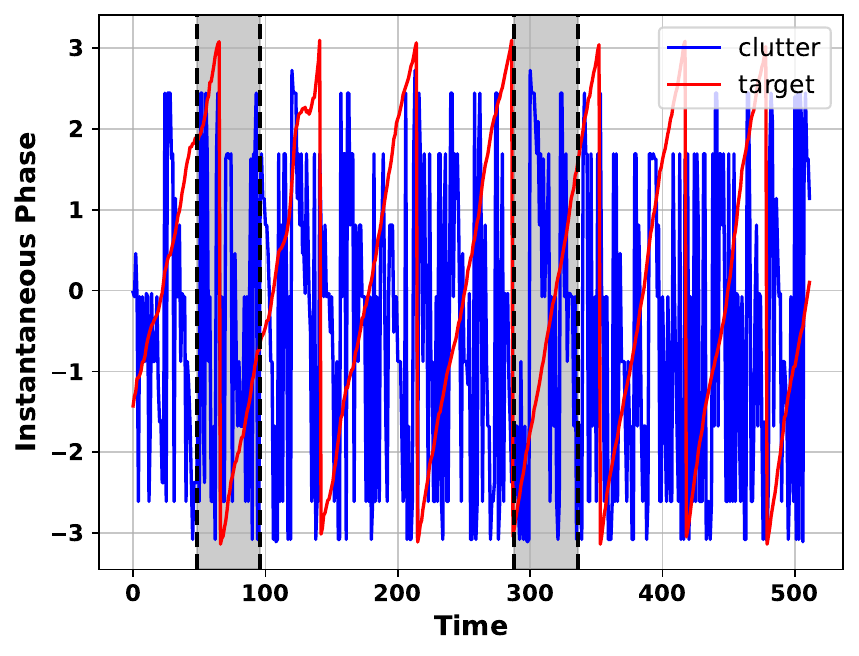} 
  \textbf{(a) IP feature}
\end{minipage}
\hfill
\begin{minipage}[b]{0.49\linewidth}
  \centering
  \includegraphics[width=\linewidth]{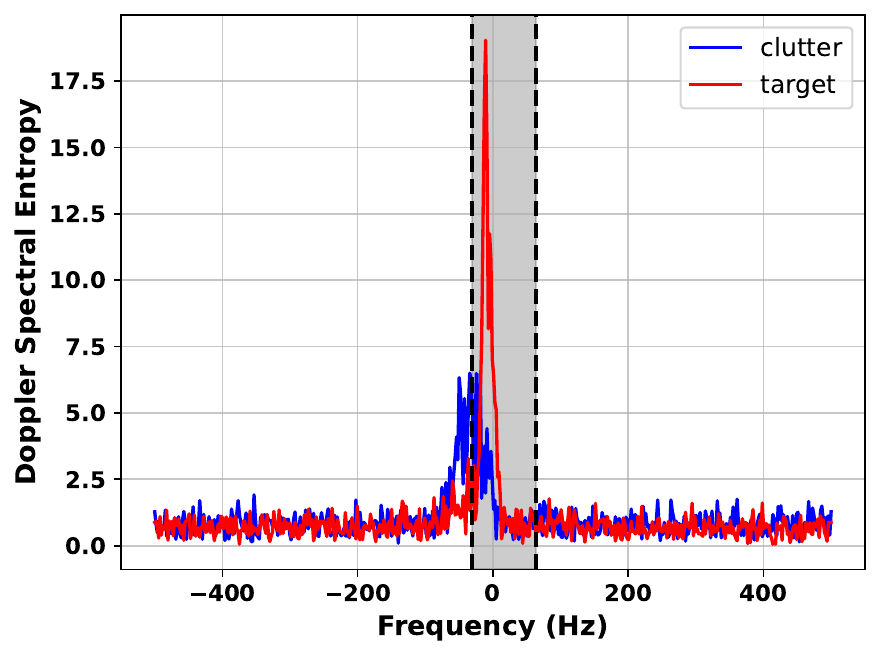} 
  \textbf{(b) DSE feature}
\end{minipage}

\vspace{0.5cm}

\begin{minipage}[b]{0.49\linewidth}
  \centering
  \includegraphics[width=\linewidth]{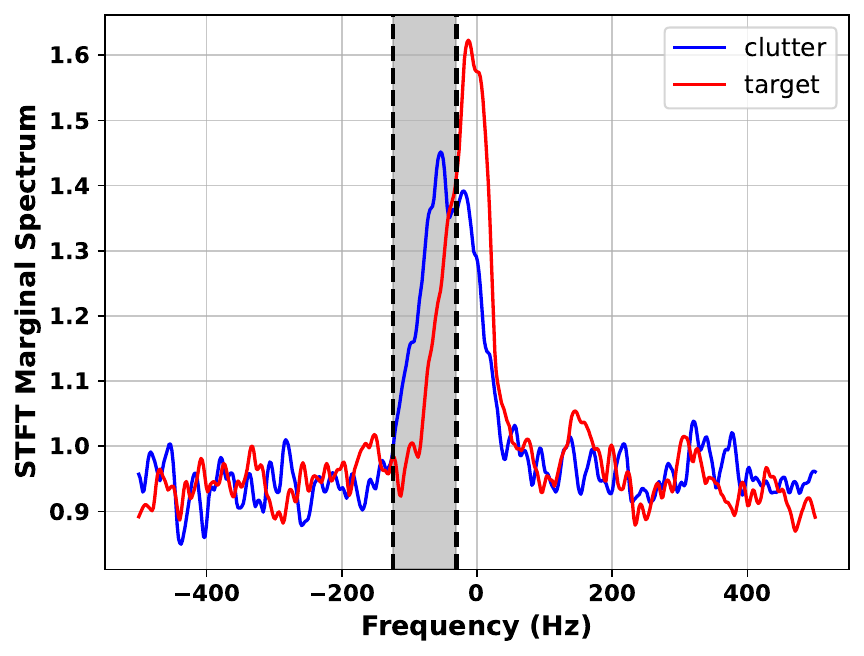} 
  \textbf{(c) SMS feature}
\end{minipage}
\hfill
\begin{minipage}[b]{0.49\linewidth}
  \centering
  \includegraphics[width=\linewidth]{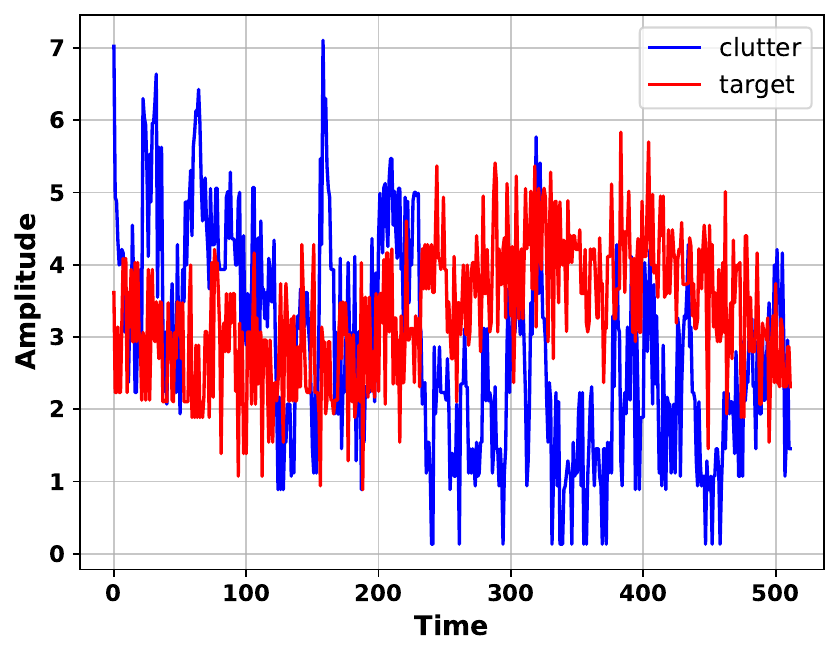} 
  \textbf{(d) Amp feature}
\end{minipage}

\vspace{0.5cm}

\begin{minipage}[b]{0.49\linewidth}
  \centering
  \includegraphics[width=\linewidth]{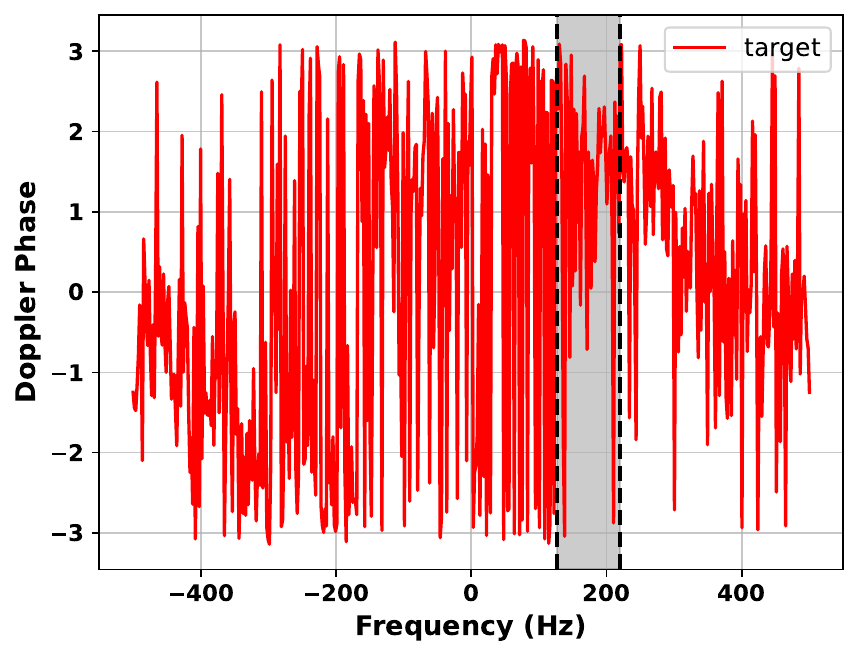} 
  \textbf{(e) DP feature (T)}
\end{minipage}
\hfill
\begin{minipage}[b]{0.49\linewidth}
  \centering
  \includegraphics[width=\linewidth]{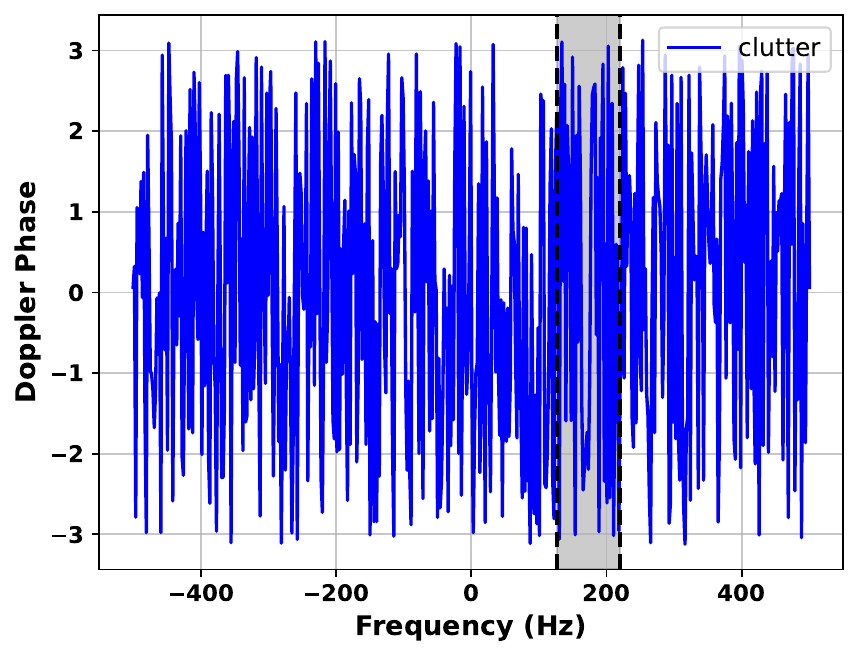} 
  \textbf{(f) DP feature (C)}
\end{minipage}

\caption{IP, DSE, SMS, Amp, and DP features for target and sea clutter echo signal on IPIX \#17 under HH polarization.}
\label{fig:res}
\end{figure}

\subsection{Significant patch selection}
In Fig.~\ref{RE}, we show our reference model architecture. A randomly initialized [CLS] token~\cite{dosovitskiy2020image} is added to the model's input. Within the Transformer, tokens interact with each other through the self-attention mechanism, defined as follows:
\begin{equation}
    \operatorname{Attention}(\boldsymbol{Q}, \boldsymbol{K}, \boldsymbol{V})=\operatorname{Softmax}\left(\frac{\boldsymbol{Q} \boldsymbol{K}^{\top}}{\sqrt{d}}\right) \boldsymbol{V},
\end{equation}
where $d$ is the length of the query vector. The [CLS] token, derived from the output of the last transformer layer, is utilized for detection. In the multi-head self-attention mechanism, the attention weights assigned by the [CLS] token to other tokens can be interpreted as indicators of their relative importance~\cite{caron2021emerging}, as the [CLS] token tends to focus on class-specific tokens while assigning less attention to those with limited useful information. To obtain a comprehensive and global measure of patch importance, we compute the average attention vector across all attention heads and all training samples in the dataset:
\begin{equation}
    \overline{\boldsymbol{a}}_{\operatorname{global}}=\frac{1}{NH}\sum_{n=1}^{N}\sum_{h=1}^{H} \boldsymbol{a}^{(n,h)},
\end{equation}
where $H$ is the total number of attention heads, $N$ is the total number of samples, and $\boldsymbol{a}^{(n,h)}$ represents the attention vector produced by the $h$-th head for the $n$-th training sample. We select only the top $y\%$ most important tokens for training, resulting in a filtered feature matrix ${F}^{\mathrm{SP}}\in R^{K^{\prime}\times L}$, where $K^{\prime}=\left\lceil\frac{Ky}{100}\right\rceil$.

\subsection{LLM for target detection}
To adapt the selected patches to the input format required by the LLM, we utilize a fully connected (FC) layer to project ${F}^{\mathrm{SP}}$ into ${F}^{\mathrm{EB}} \in \mathbb{R}^{K^{\prime} \times L^{\prime}}$. Positional encoding ${E}^{\mathrm{pos}}$ is subsequently applied to incorporate relative or absolute positional information for the patches:
\begin{equation}
    {E}^{\mathrm{pos}}_{k,2l}=\sin(\frac{k}{10000^{2l/L^{\prime}}}), \ {E}^{\mathrm{pos}}_{k,2l+1}=\cos(\frac{k}{10000^{2l/L^{\prime}}}), 
\end{equation} 
where $k$ represents the position index of the patch, and $l$ denotes the feature dimension index. The positional encoding ${E}^{\mathrm{pos}}$ is added to the embedding ${F}^{\mathrm{EB}}$ to produce ${F}^{\mathrm{PE}} = {F}^{\mathrm{EB}} + {E}^{\mathrm{pos}}$. This enriched representation ${F}^{\mathrm{PE}}$ is subsequently fed into the backbone of the LLM for further feature extraction:
\begin{equation}
{F}^{\mathrm{LLM}}=\operatorname{LLM}\left({F}^{\mathrm{PE}}\right) \in \mathbb{R}^{K^{\prime} \times L^{\prime}},
\end{equation}
where $\operatorname{LLM}(\cdot)$ represents the backbone network of the LLM. As illustrated in Fig.~\ref{RadarGPT}, to retain the universal pattern recognition capabilities of the pre-trained LLM~\cite{zhou2023one}, we fine-tune only the layer normalization layers, keeping the multi-head attention and feed-forward layers frozen. ${F}^{\mathrm{LLM}}$ are then reshaped into $\mathbb{R}^{K^{\prime}L^{\prime}}$ and passed through a FC layer followed by a softmax activation function to perform binary classification.

\section{EXPERIMENTS}
\label{sec:illust}
\subsection{Experiment setup}
We utilize nine datasets from the Intelligent PIxel Processing X-band (IPIX) database for our experiments: IPIX \#17, \#18, \#25, \#26, \#54, \#280, \#283, \#311, and \#320, all under HH polarization mode. This widely used dataset for small sea-surface target detection was collected by the IPIX radar on the east coast of Canada in November 1993. Each dataset comprises data from 14 range cells, with 131,072 samples per cell at a sampling rate of 1000 Hz. Samples from the primary cell represent target returns, while those from clutter-only cells correspond to sea clutter. Each signal sample has an observation period of 0.512 seconds.

To ensure sufficient training data, we employ overlapped segmentation following the partition rule in Eq. (\ref{Pr}), with parameters set to $M=32$ for target cells and $M=128$ for clutter cells. This process generates 4,079 target samples and over 9,000 clutter samples per dataset. The samples are divided into three groups: (1) a training set using the first 10\% of observation time for both target and clutter cells, (2) a validation set covering 10\% to 15\% of the observation time, and (3) a test set containing the remaining samples.

To control the false alarm rate, we sort the first item in the softmax output for test clutter samples in descending order. The detection threshold $\eta$ is calculated based on the desired false alarm rate $P_{fa}^{\text{d}}$ as follows: 
\begin{equation} \begin{array}{l} \eta=O_{1}(i), \ i = \lceil P_{fa}^{\text{d}} \times N_{\text{clutter}} \rceil, \end{array} \end{equation}
where $O_1$ is the first item in the sorted output array of clutter samples, $N_{\text{clutter}}$ is the number of clutter samples, and $P_{fa}^{\text{d}}$ is the expected false alarm rate, set to $P_{fa}^{\text{d}}=0.002$ in this study.

We employ a batch size of 64, the Adam optimizer~\cite{kingma2014adam}, and the cross-entropy loss function. All models are trained for 400 epochs. All experiments are conducted on a system equipped with an E5-2695v3 CPU, an NVIDIA 3090Ti GPU, and 64 GB of RAM.

\subsection{Experiment on patch selection} \label{casePS}
For patching, the size of patches is set to 48. The Transformer encoder in the reference model (RM) is configured with a model dimensionality of 128, comprising 3 layers and 16 attention heads in the multi-head self-attention mechanism. The feed-forward network (FFN) within each layer is designed with a hidden size of 256. For the pre-trained LLM, the smallest version of GPT-2 with $F = 768$ feature dimension and the first $L = 6$ layers are deployed. 

We conducted a case study on the IPIX \#17 dataset. As shown in Fig.~\ref{fig:res}, the black boxes highlight five temporal segments identified by the self-attention mechanism as most significant during training. These segments exhibit markedly higher discriminative power, validating the effectiveness of self-attention in capturing critical temporal features.

We also evaluated detection performance under different patch keep ratios. Removing less important tokens significantly improves performance, with a notable 15.5\% gain when the least important 45\% of patches are discarded, highlighting the advantage of focusing on relevant patches.

\begin{table}[h]
\centering
\begin{tabular}{c|cccc}
\hline
Patch keep ratio & \multicolumn{1}{c}{1.0} & 0.65          & 0.55           & 0.35         \\ \hline
RM        & 34.4                    & 44.9\textcolor{red}{(+10.5)}& 49.9\textcolor{red}{(+15.5)} & 47.6\textcolor{red}{(+13.2)} \\ \hline
LLM4TS          & 46.4            & 49.2\textcolor{red}{(+2.8)}  & 49.5\textcolor{red}{(+3.1)}  & 50.1\textcolor{red}{(+3.7)} \\ \hline
\end{tabular}
\caption{Detection performance on IPIX \#17 under different attentive patch keep ratio.}
\end{table}

\subsection{Experiment on detection performance}
In this section, we evaluate the proposed method against nine state-of-the-art deep learning models for marine target detection on the IPIX dataset. These models leverage the five sequence features described in Section~\ref{SF} and are categorized as follows:
\begin{enumerate}
    \item RNN-based models: RNN~\cite{wan2022sequence}, Bi-LSTM~\cite{wan2022sequence}, and GRU~\cite{cho2014learning}.
    \item Transformer-based models: Transformer~\cite{vaswani2017attention} and PatchTST~\cite{nietime}, which serves as our reference model.
    \item CNN-based models: ResNet18, ResNet34, ResNet50~\cite{xia2023target}.
    \item Hybrid models: ADN18~\cite{qu2023enhanced}, which combines time-frequency features with an enhanced CNN model.
\end{enumerate}

We further include ablation variants of our method:
\begin{enumerate}
    \item PatchTST(S): PatchTST with optimal patch retention.
    \item LLM4TS: Model with partial fine-tuning and no patch selection.
    \item LLM4TS(0): LLM4TS without pretrained transformer backbone.
    \item LLM4TS(F): LLM4TS with full fine-tuning.
    \item LLM4TS(S): Our full model with both partial fine-tuning and optimal patch selection.
\end{enumerate}

\begin{figure}[h] 
\centering
    \includegraphics[width=0.35\textwidth]{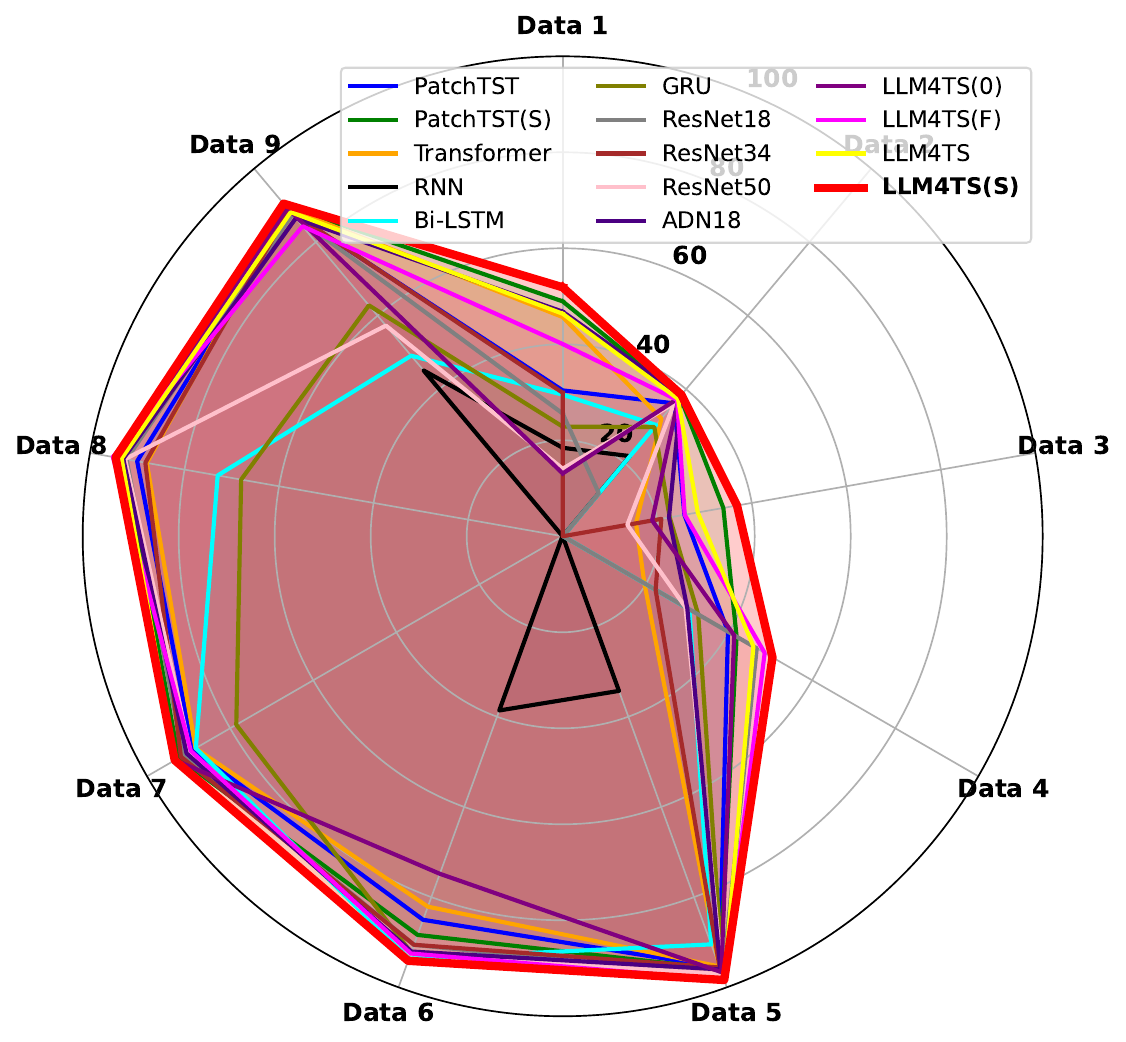}
    \caption{Model detection performance comparison on various datasets when $P_{fa}^{\text{d}}=0.002$.} \label{IPIX_result}
\end{figure}

Fig.~\ref{IPIX_result} shows detection results across nine IPIX datasets, with Table~\ref{IPIX_result1} summarizing average rates. LLM4TS(S) achieves the highest average detection rate of 72.06\%, outperforming all methods on all datasets. Key findings include:
\begin{enumerate}
    \item Partial vs. Full Fine-Tuning: LLM4TS surpasses LLM4TS(F) by 1.56\%, highlighting the efficiency of partial fine-tuning in preserving pre-trained knowledge while optimizing performance.
    \item LLM vs. Non-LLM Models: LLM4TS improves by 7.43\% over LLM4TS(0) and 5.14\% over PatchTST, highlighting the superiority of leveraging pre-trained LLM over discarding it or using non-pretrained transformer encoder.
\end{enumerate}

Moreover, optimal patch selection significantly improves both accuracy and efficiency. As shown in Table~\ref{IPIX_result1}, LLM4TS(S) and PatchTST(S) achieve detection rate improvements of 2.57\% and 4.81\%, respectively, compared to their original configurations, demonstrating the effectiveness of filtering irrelevant patches. Despite its larger network size, LLM4TS(S) processes 1334 samples per second, \textbf{\textcolor{red}{1.26}} times faster than the standard Transformer model (1074 samples per second). This efficiency stems from our patching and patch selection strategy, which significantly reduces the number of tokens processed, thereby lowering computational complexity. Additionally, the inherent inference acceleration of the GPT architecture further amplifies these gains. These advantages make LLM4TS(S) highly suitable for real-time marine target detection.

\begin{table}[h]
\centering
\begin{tabular}{|c|c|c|c|}
\hline
            & DR    & NP(M) & Throughput \\ \hline
PatchTST~\cite{nietime}    & 64.35 &  0.462/0.462 &  1092    \\ \hline
PatchTST(S) & 69.16 &   0.462/0.462    &    1438  \\ \hline
Transformer~\cite{vaswani2017attention} & 61.77 &    0.530/0.530   &  1074      \\ \hline
RNN~\cite{wan2022sequence}         & 17.60 &   0.0002/0.0002    &  2007      \\ \hline
Bi-LSTM~\cite{wan2022sequence}     & 53.87 &   0.002/0.002    &   1838     \\ \hline
GRU~\cite{cho2014learning}         & 56.32 &   0.001/0.001    &  1996      \\ \hline
ResNet18~\cite{xia2023target}    & 60.36 &  3.85/3.85   &   1053     \\ \hline
ResNet34~\cite{xia2023target}    & 58.86 &  7.22/7.22  &   1002    \\ \hline
ResNet50~\cite{xia2023target}    & 58.06 &  15.96/15.96 &   952     \\ \hline
ADN18~\cite{qu2023enhanced} & 66.18 & 14.33/14.33 & 102 \\ \hline
LLM4TS(0)      & 62.06 &  1.14/1.14     &    1128    \\ \hline
LLM4TS(F)      & 67.93 &  82.25/82.25     &    978    \\ \hline
LLM4TS      & \boldsymbol{\textcolor{blue}{69.49}} &  1.14/82.25     &      931  \\ \hline
LLM4TS(S)   & \boldsymbol{\textcolor{red}{72.06}} &  1.14/82.25     &      1334  \\ \hline
\end{tabular}
\caption{The average detection performance, network parameters (training parameters/total parameters), and interference cost per batch of different models. The best and second-best detection results are highlighted in red and blue.} \label{IPIX_result1}
\end{table}

\section{CONCLUSION}
In this paper, we propose a novel radar target detection method enhanced by LLMs. By leveraging sequence feature patching, feature patch selection, and powerful cross-modal transfer capabilities of pre-trained GPT2, we achieve significantly superior detection performance across different real-world datasets, outperforming \textbf{\textcolor{red}{nine}} other state-of-the-art models. Besides, the proposed method demonstrates acceptable inference overhead, making it suitable for practical deployment in real-world radar systems.

\section{Acknowledgement}
This work was supported by the National Natural Science Foundation of China under Grant 62388102.
\clearpage

\small
\bibliographystyle{IEEEtranN}
\bibliography{references}

\end{document}